# NEW PHYSICS AT THE INTERNATIONAL FACILITY FOR ANTIPROTON AND ION RESEARCH (FAIR) NEXT TO GSI

INGO AUGUSTIN[1,2], HANS H. GUTBROD[1,2,3], DIETER KRÄMER[1,2],
KARLHEINZ LANGANKE[1,4], HORST STÖCKER[1,3,5]

[1]*Gesellschaft für Schwerionenforschung (GSI)*
*Darmstadt, Germany*

[2]*FAIR Joint Core Team*
*Darmstadt, Germany*

[3]*Fachbereich Physik, Johann-Wolfgang-Goethe Universität*
*Frankfurt Germany*

[4]*Institute for Nuclear Physics, Technische Universität*
*Darmstadt Germany*

[5]*Frankfurt Institute for Advanced Studies (FIAS)*
*Frankfurt Germany*

The project of the international Facility for Antiproton and Ion Research (FAIR), co-located to the GSI facility in Darmstadt, has been officially started on November 7, 2007. The current plans of the facility and the planned research program will be described.

An investment of about 1 billion € will permit new physics programs in the areas of low and medium energy antiproton research, heavy ion physics complementary to LHC, as well as in nuclear structure and astrophysics. The facility will comprise about a dozen accelerators and storage rings, which will enable simultaneous operations of up to four different beams.

## 1. Introduction

In 2001, GSI, together with a large international science community, presented a Conceptual Design Report (CDR) [1] for a major new accelerator facility for beams of ions and antiprotons in Darmstadt. The concept for the new facility, now named Facility for Antiproton and Ion Research (FAIR), was based on extensive discussions and a broad range of workshops and working group reports, organized by GSI and by the international user communities over a period of several years. It also adopted priority recommendations made by high level science committees, both national and international, that have reviewed the science addressed by the proposed new facility.

About 2500 scientists and engineers from 45 countries have contributed to this effort which resulted in the FAIR Baseline Technical Report (BTR) [2]. The BTR has been accepted by the International Steering Committee as basis for the negotiations on the FAIR funding.

It is planned to found a new company for the construction and operation of FAIR in order to establish a new research facility with international ownership.

Currently 14 countries (China, Finland, France, Germany, Greece, India, Italy, Poland, Romania, Russia, Slovenia, Spain, Sweden, and the United Kingdom) have signed the Memorandum of Understanding for FAIR, indicating their wish to participate in the FAIR facility. The investment cost will be about one billion euro. For the installation activities about 2400 man years will be required. Negotiations on governmental level to secure the needed funding have started in summer 2006. The project was officially started with a ceremony on November 7, 2007. The FAIR construction plan foresees a staged completion of the facility that would allow first experimental programs to commence as early as 2014 while the entire facility would be completed in 2016.

## 2. The FAIR Facility

The experimental requirements concerning particle intensities and energies will be met by the SIS100/300 double synchrotron with a circumference of about 1,100 meters and with magnetic rigidities of 100 and 300 Tm, respectively. It constitutes the central part of the FAIR accelerator facility (see Fig. 1).

The two synchrotrons will be built on top of each other in a subterranean tunnel. They will be equipped with rapidly cycling superconducting magnets in order to minimize both construction and operating costs. For the highest intensities, the 100 Tm synchrotron will be operated at a repetition rate of 1 Hz, i.e. with ramp rates of up to 4 Tesla per second of the bending magnets. The goal of the SIS100 is to achieve intense pulsed ($5 \cdot 10^{11}$ ions per pulse) uranium beams (charge state q = 28+) at 1 GeV/u and intense ($4 \cdot 10^{13}$) pulsed proton beams at 29 GeV. For the supply of high-intensity proton beams, which are required for antiproton production, a separate proton linac as injector to the SIS18 synchrotron will be constructed.

Both heavy-ion and proton beams can be compressed down to very short bunch lengths required for the production and subsequent storage and efficient cooling of exotic nuclei (~60 ns) and antiprotons (~25 ns). These short intense ion bunches are also needed for plasma physics experiments.

Figure1. Layout of the existing (UNILAC, SIS18) and the planned FAIR facility: the superconducting synchrotrons SIS100 and SIS300, the collector ring CR, the accumulator ring

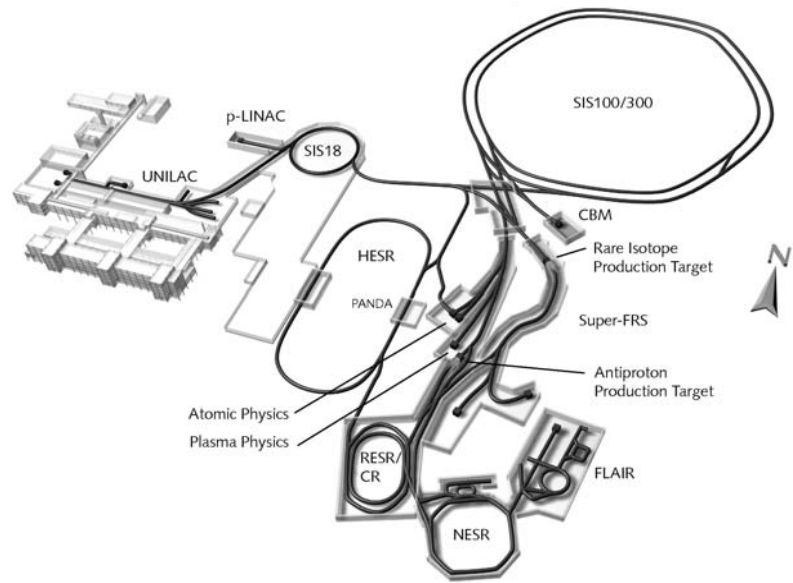

RESR, the new experimental storage ring NESR, the super fragment separator SuperFRS, the proton linac, the antiproton production target, and the high energy storage ring HESR. Also shown are the experimental stations for plasma physics, nuclear collisions (CBM), radioactive ion beams (Super-FRS), atomic physics experiments, and low-energy antiproton and atomic physics (FLAIR).

With the double ring facility, continuous beams with high average intensities of up to $3 \cdot 10^{11}$ ions per second will be provided at energies of 1 GeV/u for heavy ions, either directly from the SIS100 or by transfer to, and slow extraction from the 300 Tm ring. The SIS300 will provide high-energy ion beams of maximum energies around 45 GeV/u for $Ne^{10+}$ beams and close to 35 GeV/u for fully stripped $U^{92+}$ beams, respectively. The maximum intensities in this mode are close to $1.5 \cdot 10^{10}$ ions per spill. These high-energy beams will be extracted over time periods of 10 - 100 s in quasi-continuous mode, as the complex detector systems used for nucleus-nucleus collision experiments can accept $10^8 - 10^9$ particles per second at maximum. Slow extraction from the SIS100 is an option for extending the flexibility of parallel operation for experiments.

Adjacent to the SIS100/300 double-ring synchrotron there is a complex system of storage rings - equipped with beam cooling facilities, internal targets, and in-ring experiments - which, together with the production targets and separators for antiproton beams and radioactive secondary beams, provides an unprecedented variety of particle beams at FAIR.

From the beginning parallel operation of the different research programs was incorporated in the design of the FAIR facility. The proposed scheme of synchrotrons and storage rings, with their intrinsic cycle times for beam acceleration, accumulation, storage and cooling, respectively, has the potential to optimize a parallel and highly synergetic operation. This implies that the facility operates for the different programs more or less like a dedicated facility without reduction in luminosity, which would occur in case of simple beam splitting or steering to different experiments.

In the last seven years, substantial R&D work has been dedicated to the technological challenges. This work has been funded by the FAIR member states as well as the European Union. Considerable progress has been achieved and the principal feasibility of the proposed technical approaches been demonstrated. Prototyping of certain components has started.

## 3. Experimental Program

In most general terms, the scientific thrusts of the facility can be summarized by the following broad research goals. The first goal is to achieve a comprehensive and quantitative understanding of all aspects of matter that are governed by the strong force. Matter at the level of nuclei, nucleons, quarks and gluons is governed by the strong interaction and is often referred to as hadronic matter. The research goal of the present facility thus encompasses all aspects of hadronic matter, including the investigation of fundamental symmetries and interactions among the constituents describing the relevant degrees of freedom for this regime.

The second goal addresses many-body aspects of matter. The many-body aspects play an important and often decisive role at all levels of the hierarchical structure of matter. They govern the behavior of matter as it appears in our physical world.

These two broad science aspects, the structure and dynamics of hadronic matter and the complexity of the physical many-body system, transcend and determine the more specific research programs that will be pursued at the future facility:

- Investigations with beams of short-lived radioactive nuclei, addressing important questions about nuclei far from stability, areas of astrophysics and nucleo-synthesis in supernovae and other stellar processes, and tests of fundamental symmetries;
- The study of hadronic matter at the sub-nuclear level with beams of antiprotons, including the two key aspects: confinement of quarks and the generation of the hadron masses. They are intimately related to the existence (and spontaneous breaking) of chiral symmetry, a fundamental property of strong interactions;
- The study of compressed, dense hadronic matter in nucleus-nucleus collisions at high energies;
- The study of bulk matter in the high-density plasma state, a state of matter of interest for inertial confinement fusion and astrophysical settings;
- Studies of Quantum Electrodynamics (QED), of extremely strong (electro-magnetic) fields, and of ion-matter interactions.

The latter two aspects are the focus of the APPA experiments. Due to space limitation only the three biggest experiments can be described here.

**NuSTAR**

Secondary beams of exotic nuclei with unprecedented intensity and clarity will become available at the Super-FRagment-Separator (SFRS). This opens the unique opportunity to study the evolution of nuclear structure into the yet unexplored territory of the nuclear chart and to determine the properties of many short-lived nuclei which are produced in explosive astrophysical events and crucially influence their dynamics and associated nucleosynthesis processes.

About 700 scientists have combined to form the Nuclear Structure, Astrophysics and Reactions (NuSTAR) collaboration which will focus on experiments directly behind the production target and separator stages of the SFRS. The R3B experiment aims at exploiting the high-energy beams of short-lived nuclides to perform reaction experiments in complete kinematics with the aim to reveal new aspects of nuclear structure and collectivity in exotic nuclei. Nuclear masses and half-lives can be measured for single short-lived ions using storage ring methods developed at GSI (ILIMA). Other storage ring experiments with cooled beams concentrate on reactions with exotic nuclei (EXL) and, in a later phase, will use electrons to probe the charge distribution of such short-lived nuclei (ELISe). The spectroscopy of exotic nuclei is the focus of the HISPEC/DISPEC experiment, while LASPEC aims at the measurement of magnetic and electric moments, both planned for the low-energy branch (LEB). Penning traps at the LEB will be used for high-precision absolute mass measurements (MATS).

**PANDA**

In antiproton-proton annihilation, particles with gluonic degrees of freedom as well as particle-antiparticle pairs are copiously produced, allowing spectroscopic studies with unprecedented statistics and precision. Antiprotons of 1-15 GeV/c will therefore be an excellent tool for the following studies:

- Charmonium (cc) spectroscopy: Precision measurements of mass, width, decay branches of all charmonium states, especially for extracting information on the quark confinement. The unequaled resolution in the pp-bar formation process and small systematic uncertainties give the unique opportunity to improve dramatically our knowledge which cannot be achieved elsewhere.
- Firm establishment of the QCD-predicted gluonic excitations (charmed hybrids, glueballs) in the charmonium mass range (3-5 GeV/c$^2$) using high statistics in combination with sophisticated spin parity analysis in fully exclusive measurements.
- Search for modifications of meson properties in the nuclear medium, and their possible relationship to the partial restoration of chiral symmetry for light quarks. Particular emphasis is placed on mesons with open and hidden charm, which extends ongoing studies in the light quark sector to heavy quarks, and adds information on contributions of the gluon dynamics to hadron masses.
- Precision gamma-ray spectroscopy of single and double hyper nuclei for extracting information on their structure and on the hyperon-nucleon and hyperon-hyperon interaction.

With increasing luminosity of the HESR facility further physics opportunities will open up:

- Extraction of generalized parton distributions from pp-bar annihilation.
- D-meson decay spectroscopy (rare leptonic and hadronic decays).
- Search for CP violation in the charm and strangeness sector (D meson decays, system).

**CBM**

The CBM Collaboration proposes to build a dedicated heavy-ion experiment to investigate the properties of highly compressed baryonic matter as it is produced in nucleus-nucleus collisions at the future accelerator facility in Darmstadt. The goal is to explore the QCD phase diagram in the region of moderate temperatures but very high baryon densities. The envisaged research program includes the study of key questions of QCD like confinement, chiral symmetry restoration and the nuclear equation of state at high densities. The most promising diagnostic probes are vector mesons decaying into di-lepton pairs, strangeness and charm. It is intended to perform comprehensive measurements of hadrons, electrons and photons created in collisions of heavy nuclei.

CBM will be a fixed target experiment, which covers a large fraction of the populated phase space. The major experimental challenge is posed by the extremely high reaction rates of up to $10^7$ events/second. These conditions require unprecedented detector performances concerning speed and radiation hardness. The detector layout comprises a high resolution Silicon Tracking System in a magnetic dipole field for particle momentum and vertex determination, Ring Imaging Cherenkov Detectors and Transition Radiation Detectors for the identification of electrons, an array of Resistive Plate Chambers for hadron identification via TOF measurements, and an electromagnetic calorimeter for the identification of electrons, photons and muons. The detector signals are processed by a high-speed data acquisition and trigger system.